\documentstyle[aps]{revtex}
\input epsf
\begin{document}

\twocolumn[\hsize\textwidth\columnwidth\hsize\csname
@twocolumnfalse\endcsname

\title{Monte Carlo energy and variance minimization techniques for
optimizing many--body wave functions}

\author{P.~R.~C.~Kent, R.~J.~Needs, and G.~Rajagopal}

\address{Cavendish Laboratory, Madingley Road, Cambridge CB3 0HE, UK}

\date{\today}
 
\maketitle

\begin{abstract}
\begin{quote}
\parbox{16 cm}{\small 
We investigate Monte Carlo energy and variance minimization techniques
for optimizing many--body wave functions.  Several variants of the
basic techniques are studied, including limiting the variations in the
weighting factors which arise in correlated sampling estimations of
the energy and its variance.  We investigate the numerical stability
of the techniques and identify two reasons why variance minimization
exhibits superior numerical stability to energy minimization.  The
characteristics of each method are studied using a non--interacting
64--electron model of crystalline silicon.  While our main interest is
in solid state systems, the issues investigated are relevant to Monte
Carlo studies of atoms, molecules and solids.  We identify a robust
and efficient variance minimization scheme for optimizing wave
functions for large systems.}
\end{quote}
\end{abstract}

\pacs{PACS: 71.15.-m, 02.70.Lq}

]
\narrowtext

\section{Introduction}
\label{introduction}

Accurate approximations to many--body wave functions are crucial for
the success of quantum Monte Carlo (QMC) calculations.  In the
variational quantum Monte Carlo (VMC) method~\cite{vmc,hammond}
expectation values are calculated as integrals over configuration
space, which are evaluated using standard Monte Carlo techniques.  In
the more sophisticated diffusion Monte Carlo (DMC)
method~\cite{hammond,dmc} imaginary time evolution of the
Schr\"{o}dinger equation is used to calculate very accurate
expectation values.  Importance sampling is included via a trial wave
function and the fermion sign--problem is evaded by using the
fixed--node approximation.

The most costly part of VMC and DMC calculations is normally the
evaluation of the trial wave function (and its gradient and Laplacian)
at many different points in configuration space.  The accuracy of the
trial wave function controls the statistical efficiency of the
algorithm and limits the final accuracy that can be obtained.  It is
therefore necessary to use trial wave functions which are as accurate
as possible yet can be computed rapidly.  By far the most common type
of trial wave function used in VMC and DMC calculations for atoms,
molecules and solids is the Slater--Jastrow form:

\begin{equation}
\Phi = \sum_n \beta_n D^{\uparrow}_n D^{\downarrow}_n \exp \left[
-\sum_{i>j}^{N}u({\bf r}_i,{\bf r}_j)+\sum_{i}^{N}\chi ({\bf
r}_i)\right] \;\;,
\label{Phi}
\end{equation}

\noindent where $N$ is the number of electrons, $D^{\uparrow}_n$ and
$D^{\downarrow}_n$ are Slater determinants of spin--up and spin--down
single--particle orbitals, the $\beta_n$ are coefficients, $\chi$ is a
one--body function, and $u$ is a relative--spin--dependent two--body
correlation factor.

The functions $u$ and $\chi$ normally contain variable parameters, and
one may also wish to vary the $\beta_n$ and parameters in the
single--particle orbitals forming the Slater determinants.  The values
of the parameters are obtained via an optimization procedure.  Typical
solid state problems currently involve optimizing of order 10$^2$
parameters for 10$^3$--dimensional functions.  These optimization
problems are delicate and require careful handling.

In this paper we investigate several variants of energy and variance
minimization techniques.  Our aims are (i) to identify the reasons why
variance minimization exhibits superior numerical stability to energy
minimization, and (ii) to identify the best variance minimization
scheme for optimizing wave functions in large systems.  We concentrate
on two areas, the nature of the objective function
(Section~\ref{objective}) and the effects of approximating the
required integrals by finite sums (Section~\ref{finite}).  In
Section~\ref{tests} we use a 64--electron model of crystalline silicon
to investigate the behaviour of various optimization schemes, while in
Section~\ref{conclusions} we draw our conclusions.

\section{The Objective Function}
\label{objective}

In order to optimize a wave function we require an objective function,
i.e., a quantity which is to be minimized with respect to a set of
parameters, $\{\alpha\}$.  The criteria that a successful objective
function should satisfy for use in a Monte Carlo optimization
procedure are that (i) the global minimum of the objective function
should correspond to a high quality wave function, (ii) the variance
of the objective function should be as small as possible, and (iii)
the minimum in the objective function should be as sharp and deep as
possible.  One natural objective function is the expectation value of
the energy,

\begin{equation}
E_{{\rm V}} = \frac{\int \Phi^2(\alpha) \; [\Phi^{-1}(\alpha) \hat{H}
\Phi(\alpha)]\,d{\bf R}}{\int \Phi^2(\alpha) \,d{\bf R}} \;\;,
\label{E_V}
\end{equation}

\noindent where the integrals are over the 3$N$--dimensional
configuration space.  The numerator is the integral over the
probability distribution, $\Phi^2(\alpha)$, of the local energy,
$E_{{\rm L}}(\alpha) = \Phi^{-1}(\alpha) \hat{H} \Phi(\alpha)$.

In fact the energy is not the preferred objective function for wave
function optimization, and the general consensus is that a better
procedure is to minimize the variance of the energy, which is given by

\begin{equation}
A(\alpha) = \frac{\int \Phi^2(\alpha) \; [E_{{\rm L}}(\alpha) -
E_{{\rm V}}(\alpha)]^2 \, d{\bf R}}{\int \Phi^2(\alpha) \, d{\bf R}}
\;\;.
\label{A_1}
\end{equation}

Optimizing wave functions by minimizing the variance of the energy is
actually a very old idea, having being used in the 1930's.  The first
application using Monte Carlo techniques to evaluate the integrals
appears to have been by Conroy~\cite{conroy}, but the present
popularity of the method derives from the developments of Umrigar and
coworkers~\cite{uww,umrigar}.  A number of reasons have been advanced
for preferring variance minimization, including: (i) it has a known
lower bound of zero, (ii) the resulting wave functions give good
estimates for a range of properties, not just the energy, (iii) it can
be applied to excited states, (iv) efficient algorithms are known for
minimizing objective functions which can be written as a sum of
squares, and (v) it exhibits greater numerical stability than energy
minimization.  The latter point is very significant for applications
to large systems.

The minimum possible value of $A(\alpha)$ is zero. This value is
obtained if and only if $\Phi(\alpha)$ is an exact eigenstate of
$\hat{H}$.  Minimization of $A(\alpha)$ has normally been carried out
via a correlated sampling approach in which a set of configurations
distributed according to $\Phi^2(\alpha_0)$ is generated, where
$\alpha_0$ is an initial set of parameter values.  $A(\alpha)$ is then
evaluated as

\begin{equation}
A(\alpha) = \frac{\int \Phi^2(\alpha_0) \;
w(\alpha)\; [E_{{\rm L}}(\alpha) -
E_{{\rm V}}(\alpha)]^2 \, d{\bf R}}{\int \Phi^2(\alpha_0) \;
w(\alpha) \, d{\bf R}} \;\;,
\label{A_2}
\end{equation}

\noindent where the integrals contain a weighting factor,
$w(\alpha)$, given by

\begin{equation}
w(\alpha) = \frac{\Phi^2(\alpha)}{\Phi^2(\alpha_0)} \;\;.
\label{W}
\end{equation}

\noindent $A(\alpha)$ is then minimized with respect to the parameters
$\{\alpha\}$.  The set of configurations is normally regenerated
several times with the updated parameter values so that when
convergence is obtained $\{\alpha_0\}$=$\{\alpha\}$.  A variant of
Eq.~\ref{A_2} is obtained by replacing the energy $E_{{\rm
V}}(\alpha)$ by a fixed value, $\bar{E}$, giving

\begin{equation}
B(\alpha) = \frac{\int \Phi^2(\alpha_0) \;
w(\alpha)\; [E_{{\rm L}}(\alpha) -
\bar{E}]^2 \, d{\bf R}}{\int \Phi^2(\alpha_0) \;
w(\alpha) \, d{\bf R}} \;\;.
\label{B}
\end{equation}

\noindent Note that if $\bar{E} \leq E_0$, where $E_0$ is the exact
ground state energy, then the minimum possible value of $B(\alpha)$
occurs when $\Phi$=$\Phi_0$, the exact ground state wave function.
Minimization of $B(\alpha)$ is equivalent to minimizing a linear
combination of $E_{\rm V}$ and $A(\alpha)$.  The absolute minima of
both $E_{\rm V}$ and $A(\alpha)$ occur when $\Phi$=$\Phi_0$. If both
of the coefficients of $E_{\rm V}$ and $A(\alpha)$ in the linear
combination are positive, which is guaranteed if $\bar{E} \leq E_0$,
then it follows that the absolute minimum of $B(\alpha)$ occurs at
$\Phi$=$\Phi_0$.  Using this method with $\bar{E} \leq E_0$ allows
optimization only of the ground state wave function.

Although minimization of $A(\alpha)$ or $B(\alpha)$ using correlated
sampling methods has often been successful, in some cases the
procedure can exhibit a numerical instability.  Two situations where
this is likely to occur have been identified.  The first is when the
nodes of the trial wave function are allowed to alter during the
optimization process.  A similar instability can arise when the number
of electrons in the systems becomes large, which can result in an
instability even if the nodes of the trial wave function remain fixed.
The characteristic of these numerical instabilities is that during the
minimization procedure a few configurations (often only one) acquire a
very large weight.  The estimate of the variance is then reduced
almost to zero by a set of parameters which are found to give
extremely poor results in a subsequent QMC calculation.  When the
nodes of the trial wave function are altered large weights are most
likely to occur for configurations close to the zeros of the
probability distribution $\Phi^2(\alpha_0)$.  Large weights can also
occur when varying the Jastrow factor if the number of electrons, $N$,
is large.  For a small change in the one--body function, $\delta
\chi$, the local energy changes by an amount {\it proportional} to $N
\delta \chi$, but the weight is multiplied by a factor which is {\it
exponential} in $N \delta \chi$, which can result in very large or
very small weights if $N$ is large.  A similar argument holds for
changes in the two--body term, which shows an even more severe
potential instability because the change in the two--body term scales
like $N^2$.

The instability due to the weights has been noticed by many
researchers.  In principle one could overcome this instability by
using more configurations, but the number required is normally
impossibly large.  Various practical ways of dealing with this
instability have been devised.  One method is to limit the upper value
of the weights~\cite{filippi} or to set the weights equal to
unity~\cite{schmidt,varmin_96}.  Schmidt and Moskowitz~\cite{schmidt}
set the weights equal to unity in calculations for small systems in
which the nodes were altered.  An alternative approach is to draw the
configurations from a modified probability distribution which is
positive definite, so that the weights do not get very
large.~\cite{barnett} In our calculations for large systems of up to
1000 electrons~\cite{finsize_98} we also set the weights equal to
unity while optimizing the Jastrow factor.  When using the correlated
sampling approach, whether or not the weights are modified, better
results are obtained by periodically regenerating a new set of
configurations chosen from the distribution $\Phi^2(\alpha)$, where
$\{\alpha\}$ is the updated parameter set.  This helps the convergence
of the minimization procedure.  One can also restrict the allowed
variation in the parameters $\{\alpha\}$ before regenerating a new set
of configurations, but this can slow the convergence.  We
found~\cite{varmin_96} that setting the weights to unity allowed us to
alter the parameters by a larger amount before we had to regenerate
the configurations with the new set of parameters.  After a few
(typically three or four) regenerations we found that the parameters
had converged to stable values giving a small variance and low energy
in a subsequent VMC calculation.

These strategies can often overcome the numerical instability.  Our
goal is to apply QMC methods to large systems with many inequivalent
atoms, which will require wave functions for many electrons with many
variable parameters.  We would like to be able to optimize the
determinantal part of the wave function as well as the Jastrow factor,
which has only recently been attempted for solids~\cite{fahy_opt}, and
we would also like to optimize excited states as well as ground
states.  In order to accomplish these goals we will need to improve
our optimization techniques.  In this paper we analyse energy and
variance minimization techniques, in the expectation that a deeper
understanding of the issues of numerical stability will lead to
improved algorithms.

First we analyse the procedure of setting the weights to unity, which
gives a new objective function, $C(\alpha)$, where

\begin{equation}
C(\alpha) = \frac{\int \Phi^2(\alpha_0) \; [E_{{\rm L}}(\alpha) -
E_{C}(\alpha)]^2 \, d{\bf R}}{\int \Phi^2(\alpha_0) \, d{\bf R}}
\;\;,
\label{C}
\end{equation}

\noindent and

\begin{equation}
E_{C} = \frac{\int \Phi^2(\alpha_0) \; [\Phi^{-1}(\alpha) \hat{H}
\Phi(\alpha)] \, d{\bf R}}{\int \Phi^2(\alpha_0) \, d{\bf R}} \;\;,
\label{E_C}
\end{equation}

\noindent is the unweighted energy.  The objective function
$C(\alpha)$ has the property that its absolute minimum is zero and
that this value is obtained if and only if $\Phi(\alpha)$ is an exact
eigenstate of $\hat{H}$, because for an exact eigenstate $E_{{\rm L}}
= E_C$.  The absolute minima of $C(\alpha)$ are therefore at the same
positions as those of $A(\alpha)$ and therefore $C(\alpha)$ should be
a satisfactory objective function.  As we will show by explicit
example in Section~\ref{tests}, the advantage of $C(\alpha)$ is that
it has a lower variance than $A(\alpha)$, especially when $\alpha_0$
and $\alpha$ differ significantly.  A similar analysis can be applied
to the case where the weights are subject to an upper limit, and we
will refer to all such expressions with modified weights as variants
of $C$ and $E_C$.

The objective function $C(\alpha)$ contains the unweighted energy
$E_C$.  As we will show by explicit example in Section~\ref{tests},
the ground state of $\hat{H}$ does not necessarily correspond to the
minimum value of $E_C$.  The energy $E_C$ is therefore not a
satisfactory objective function in its own right.  If we replace the
energy $E_{C}(\alpha)$ in Eq.~\ref{C} by some other energy $\bar{E}$,
then the minima of the objective function occur at the eigenstates of
$\hat{H}$ if and only if $\bar{E}$ evaluated with the exact wave
function is equal to the exact energy of the eigenstate.  This
requirement still allows freedom in the choice of $\bar{E}$, and the
following form is sufficient,

\begin{equation}
\bar{E} = \frac{\int p({\bf R}) \; E_{{\rm L}}(\alpha) \,
d{\bf R}}{\int p({\bf R}) \, d{\bf R}} \;\;,
\label{barE}
\end{equation}

\noindent where $p({\bf R})$ is any probability distribution.  This
demonstrates that we can alter the weights in the energy $E_C$ and the
variance $C$ independently, without shifting the positions of the
absolute minima of $C$.  In this work we have not investigated this
freedom and we have always used the same weights for $E_C$ and $C$.

The above analysis applies for wave functions with sufficient
variational freedom to encompass the exact wave function.  In
practical situations we are unable to find exact wave functions and it
is important to consider the effect this has on the optimization
process.  Although the objective functions $A(\alpha)$ and $C(\alpha)$
are unbiased in the sense that the exact ground state wave function
corresponds to an absolute minimum, $C(\alpha)$ is biased in the sense
that for a wave function which cannot be exact the optimized
parameters will not exactly minimize the true variance.  We refer to
this as a ``weak bias'' because it disappears as the wave function
tends to the exact one.  In practice this is not a problem because in
minimizing $C(\alpha)$ we regenerate the configurations several times
with the updated distribution until convergence is obtained, so that
minimization of $A(\alpha)$ and $C(\alpha)$ turns out to give almost
identical parameter values.  On the other hand, the unweighted energy,
$E_C$, shows a ``strong bias'' in the sense that the nature of its
stationary points are very different from those of the properly
weighted energy.  The ability to alter the weights while not affecting
the positions of the minima is an important advantage of variance
minimization over energy minimization, which we believe is one of the
factors which leads to the greater numerical stability of variance
minimization.

\section{Further Effects of Finite Sampling}
\label{finite}

In the previous section we described the numerical instability arising
from the weighting factors.  The origin of this problem lies in
approximating the integrals by the average of the integrand over a
finite set of points in configuration space.  There is another
important issue connected with the approximation of finite sampling,
which is whether the positions of the minima of the objective function
are altered by the finite sampling itself.

Consider the objective function $A(\alpha)$, in the case where the
trial wave function has sufficient variational freedom to encompass
the exact wave function.  Approximating Eq.~\ref{A_2} by an average
over the set $\{{\bf R}_i\}$ containing $N_s$ configurations drawn
from the distribution $\Phi^2(\alpha_0)$ gives

\begin{equation}
A^{N_s} = \frac{\sum_i^{N_s} w({\bf R}_i;\alpha) [E_{{\rm L}}({\bf
R}_i;\alpha) - E_{{\rm V}}(\{{\bf R}_i\};\alpha)]^2} {\sum_i^{N_s}
w({\bf R}_i;\alpha)} \;\;.
\label{A^N_s}
\end{equation}

\noindent The eigenstates of $\hat{H}$ give $A^{N_s} = 0$ for {\it
any} size of sample because $E_{{\rm L}} = E_{{\rm V}}$ for an
eigenstate.  Clearly this result also holds for $C(\alpha)$.  This
behaviour contrasts with that of the variational energy, $E_{{\rm
V}}$.  Consider a finite sampling of the variational energy of
Eq.~\ref{E_V}, where the configurations are distributed according to
$\Phi^2(\alpha_0)$ and properly weighted,

\begin{equation}
E_{{\rm V}}^{N_s} = \frac{\sum_i^{N_s}w({\bf R}_i;\alpha) E_{{\rm
L}}({\bf R}_i;\alpha)} {\sum_i^{N_s} w({\bf R}_i;\alpha)} \;\;.
\label{E_V^N_s}
\end{equation}

\noindent The global minima of $E_{{\rm V}}^{N_s}$ are not guaranteed
to lie at the eigenstates of $\hat{H}$ for a finite sample.  The fact
that the positions of the global minima of $A(\alpha)$ and $C(\alpha)$
are robust to finite sampling is a second important advantage of
variance minimization over energy minimization.

\section{Tests of minimization procedures}
\label{tests}
 
We now investigate the performance of the various energy and variance
minimization techniques for a solid state system.  We would like to
know the exact wave function for our test system, and therefore we
have chosen a non--interacting system.  We model the valence states
of silicon in the diamond structure, using periodic boundary
conditions to simulate the solid.  The fcc simulation cell contains 16
atoms and 64--electrons.  The electrons are subject to a local
potential which is described by two Fourier components, $V_{111}$ =
-0.1 a.u., and $V_{220}$ = -0.06 a.u., chosen to give a reasonable
description of the valence bandstructure of silicon.  The value of
$V_{111}$ is in good agreement with empirical pseudopotential form
factors for silicon~\cite{cohen}, while the value of $V_{220}$ is
somewhat larger.  Overall this model gives a reasonable description of
the valence states of silicon and retains the essential features for
testing the optimization techniques.

The ``exact'' single--particle orbitals were obtained by diagonalizing
the Hamiltonian in a plane--wave basis set containing all waves up to
an energy cutoff of 15 a.u.  This basis set is still incomplete, but
the square root of the variance of the energy is about 0.02 eV per
atom, which is negligible for our purposes.  We have added a
variational parameter, $\alpha$, in the form of a $\chi$ function with
the full symmetry of the diamond structure:

\begin{equation}
\chi ({\bf r})=\alpha\left( \mathop{\displaystyle \sum } \limits_{{\bf
G}}P_{{\bf G}}e^{{\rm i}{\bf G.r}}\right) \;\;,
\end{equation}

\noindent where ${\bf G}$ labels the 8 reciprocal lattice vectors of
the [111] star and $P_{{\bf G}}$ is a phase factor associated with the
non--symmorphic symmetry operations.  The exact value of the
parameter, $\alpha$, is, of course, zero.  To model the situation
where the wave function does not possess the variational freedom to
encompass the exact one we used a smaller basis set cutoff of 2.5 a.u.
The variational energy from this wave function is 0.35 eV per atom
above the exact value, which is typical of the values we encounter in
our solid state calculations.  The optimal value of $\alpha$ for this
inexact wave function is very close to zero. 

This model exhibits all the numerical problems we have encountered in
optimization procedures.  In practical situations one may have more
electrons and more parameters to optimize, which makes the numerical
instabilities more pronounced.  In order to analyse the behaviour in
detail we have evaluated the variance of the objective functions.  We
found that unfeasibly large numbers of electron configurations were
required to obtain accurate values of the variance of the objective
functions for wave functions with many more electrons and variables
parameters than used in our model system.  We stress that when the
numerical instabilities are more pronounced it is even more
advantageous to adopt the optimization strategies recommended here.

We generated samples of $0.96\times 10^{6}$ statistically independent
electronic configurations which were used to calculate the quantities
involved in the various optimization schemes.  In practical
applications, a typical number of configurations used might be $10^4$,
but we found it necessary to use a much larger number to obtain
sufficiently accurate values of the different objective functions and,
particularly, their variances.  In a practical application an
objective function, say $C(\alpha)$, is evaluated using, say, $10^4$
configurations.  The quantities of interest are then $C(\alpha)$ and
its variance calculated as averages over blocks of $10^4$
configurations.  Because the numerator in $C(\alpha)$ contains $E_{\rm
V}$, which is itself a sum over configurations, the values of
$C(\alpha)$ and its variance depend on the number of configurations in
the block.  The variance of $C(\alpha)$ calculated as such a block
average is much more sensitive to the block size than the value of
$C(\alpha)$.  As the number of configurations in the block increases
the values of $C(\alpha)$ and its variance converge to their true
values. (Analagous arguments hold for $A(\alpha)$.)  Quoting all our
results as a function of the block size would result in an enormous
increase in the amount of data.  However, for our silicon model, the
variances of the objective functions are close to their true
asymptotic values for block sizes of $10^4$ configurations or greater,
so the values at the limit of large block sizes are the relevant ones
for practical applications, and these are the values we quote here.

The configurations were generated by a Metropolis walk distributed
according to $\Phi^2$, using the inexact reduced basis--set wave
function.  An optimization procedure typically starts with
non--optimal parameter values which are improved during the
optimization procedure.  We present results for configurations
generated with the non--optimal value of $\alpha_0$ = 0.03, which
gives results typical of the starting value for an optimization, and
$\alpha_0$ = 0, which is the final value from a successful
optimization procedure.  The qualitative behaviour is not strongly
influenced by the value of $\alpha_0$.

\begin{figure}
\epsfysize=8cm \epsfbox{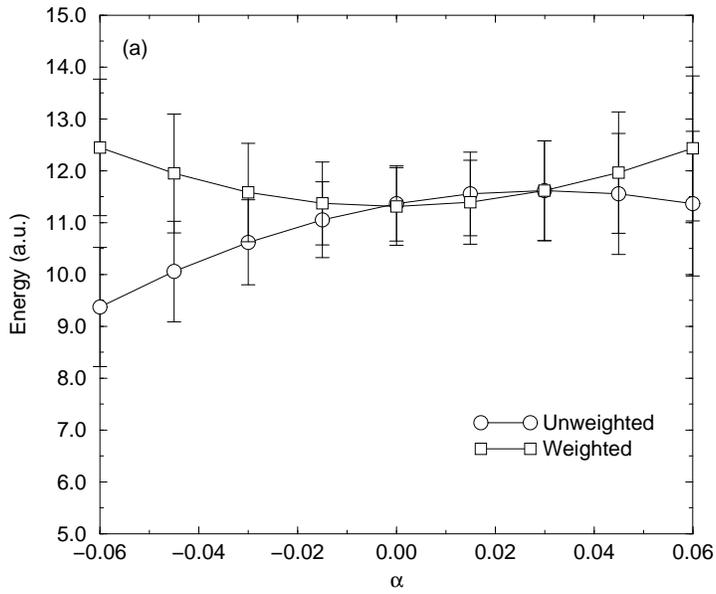}

\epsfysize=8cm \epsfbox{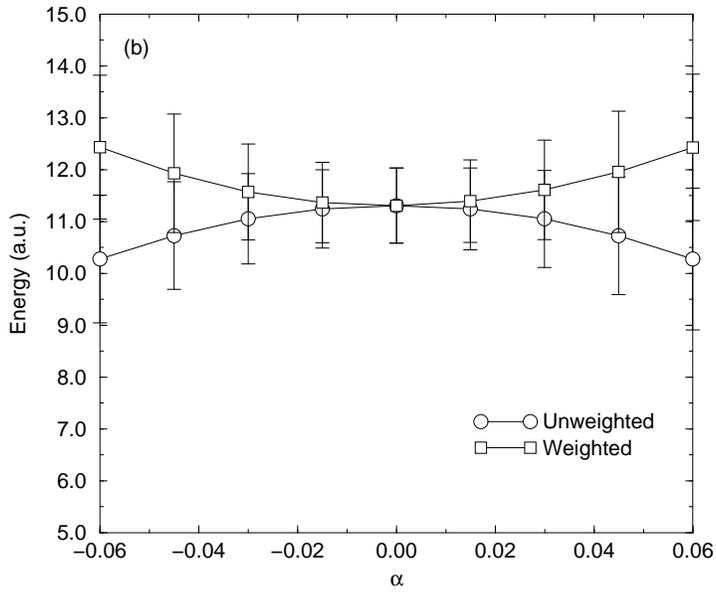}
\caption{Weighted and unweighted mean energies and standard
deviations, shown as error bars, for configurations generated with (a)
$\alpha_0$=0.03 and (b) $\alpha_0$=0.}
\label{fig1}
\end{figure}

First we consider energy minimization.  In Fig.~\ref{fig1} we plot the
weighted and unweighted mean energies, $E_{\rm V}$ and $E_C$, and
their variances as a function of $\alpha$, with configurations
generated from $\alpha_0$ = 0.03 (Fig.~\ref{fig1}a) and $\alpha_0$ = 0
(Fig.~\ref{fig1}b).  The unweighted mean energy has a $\it maximum$ at
$\alpha_0$, i.e., the value from which the configurations were
generated.  This result can be understood as follows.  Consider a wave
function of the form

\begin{equation}
\Phi = \sum_n \beta_n D^{\uparrow}_n D^{\downarrow}_n \exp
\left[\sum_k \alpha_k J_k \right] \;\;,
\label{Phi_linear}
\end{equation} 

\noindent where the $\alpha_k$ are parameters, and the $J_k$ are
correlation functions.  The mean unweighted energy can be written as

\begin{equation}
E_{C}(\{\alpha_k\}) = (\alpha_k - \alpha_{k0})\,G_{kl}\,(\alpha_l -
\alpha_{l0})\;\;+\;\;{\rm constant}\;\;,
\label{E_unweight}
\end{equation}

\noindent where $\alpha_{k0}$ are the parameter values from which the
configurations are generated and 

\begin{equation}
G_{kl} = \frac{-\frac{1}{2} \int \Phi^2(\{\alpha_{k0}\}) \sum_i
\nabla_i J_k \nabla_i J_l \, d{\bf R}}{\int \Phi^2(\{\alpha_{k0}\}) \,
d{\bf R}}\;\;.
\end{equation}

\noindent The $G_{kl}$ and the constant term depend on the
$\alpha_{k0}$ but not on the $\alpha_k$.  When there is only a single
parameter, $G$ is negative, so that $E_C$ is a quadratic function with
a maximum at $\alpha_k = \alpha_{k0}$.  When there is more than one
parameter the stationary point of the quadratic can be a maximum,
minimum or saddle point, which is not acceptable behaviour for an
objective function.  The weights may be altered in other ways, such as
limiting their upper value, but if the weights are altered the minima
of the energy are moved, which is a ``strong bias'' in the objective
function.  If one insists on using an energy minimization method,
weighting $\it must$ be used.

\begin{figure}
\epsfysize=8cm \epsfbox{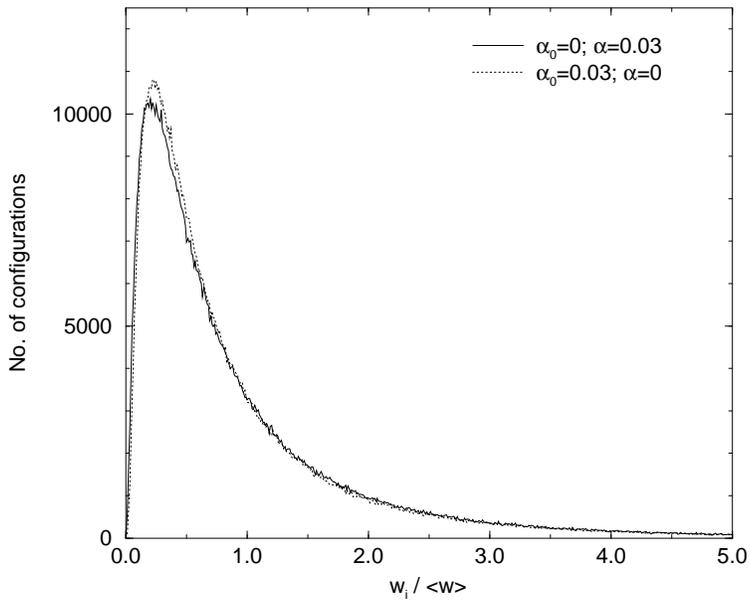}
\caption{Distributions of weights for configurations generated with
$\alpha_0$=0.03 and $\alpha_0$=0, evaluated with $\alpha$=0 and
$\alpha$=0.03, respectively.}
\label{fig2}
\end{figure}

\begin{figure}
\epsfysize=8cm \epsfbox{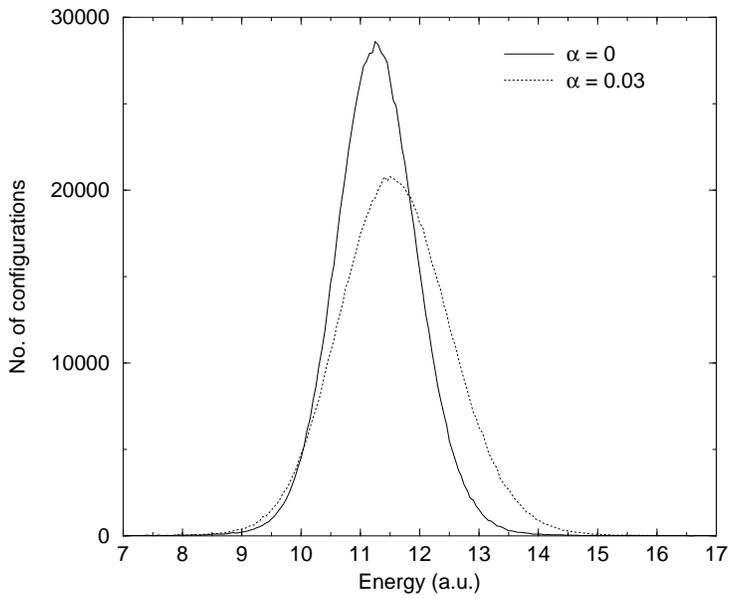}
\caption{Distributions of the local energies for $\alpha$=0 and
$\alpha$=0.03.}
\label{fig3}
\end{figure}

We now investigate the distributions of the weights and the local
energies.  In Fig.~\ref{fig2} we plot the distributions of the weights
for $\alpha_0$ = 0.03 and $\alpha$ = 0 and for $\alpha_0$ = 0 and
$\alpha$ = 0.03, while in Fig.~\ref{fig3} we plot the corresponding
distributions of the local energies.  The distributions of the weights
resemble Poisson distributions, but the square roots of the variances
are significantly greater than the means, so there are more
configurations at large weights than for a Poisson distribution with
the same mean.  The local energies follow normal distributions
relatively well.  As expected, the distributions of the local energies
is wider for the $\alpha_0$ = 0.03 wave function.  Closer inspection
reveals that the distribution of the local energies is not exactly
normal because the actual distributions have ``fat tails''.  The
outlying energies result from outlying values of the kinetic energy.
The standard deviations are $\sigma$ = 0.964 a.u. and $\sigma$ = 0.726
a.u., for $\alpha_0$ = 0.03 and 0, respectively.  The expected
percentage of configurations beyond 3$\sigma$ from the mean of a
normal distribution is 0.27\%, but the actual percentages are 0.443\%
and 0.608\% for $\alpha_0$ = 0.03 and 0, respectively.  Although these
outlying local energies give a negligible contribution to the mean
energy, calculated with or without weighting, and only a very small
contribution to the values of the variance--like objective functions,
$A(\alpha)$, $B(\alpha)$, and $C(\alpha)$, they give significant
contributions to the variances of the variance--like objective
functions.

It is highly undesirable for an objective function to have a large
variance.  A larger variance implies that a greater number of
configurations is required to determine the objective function to a
given accuracy.  However, as noted above, only the variances and not
the means of $A(\alpha)$, $B(\alpha)$, and $C(\alpha)$ are
significantly affected by these outlying configurations.  We therefore
limit the outlying local energies.  An alternative would be to delete
the outlying configurations, but this introduces a greater bias and is
not as convenient in correlated--sampling schemes.  The limiting must
be done by the introduction of an arbitrary criterion, which we have
implemented as follows.  First we calculate the standard deviation of
the sampled local energies, $\sigma$.  We then calculate limiting
values for the local energy as those beyond which the total expected
number of configurations based on a normal distribution is less than
$\Delta$, where

\begin{equation}
\Delta = N_s \times 10^{-p}\;\;, 
\label{Delta}
\end{equation}

\begin{figure}
\epsfysize=8cm \epsfbox{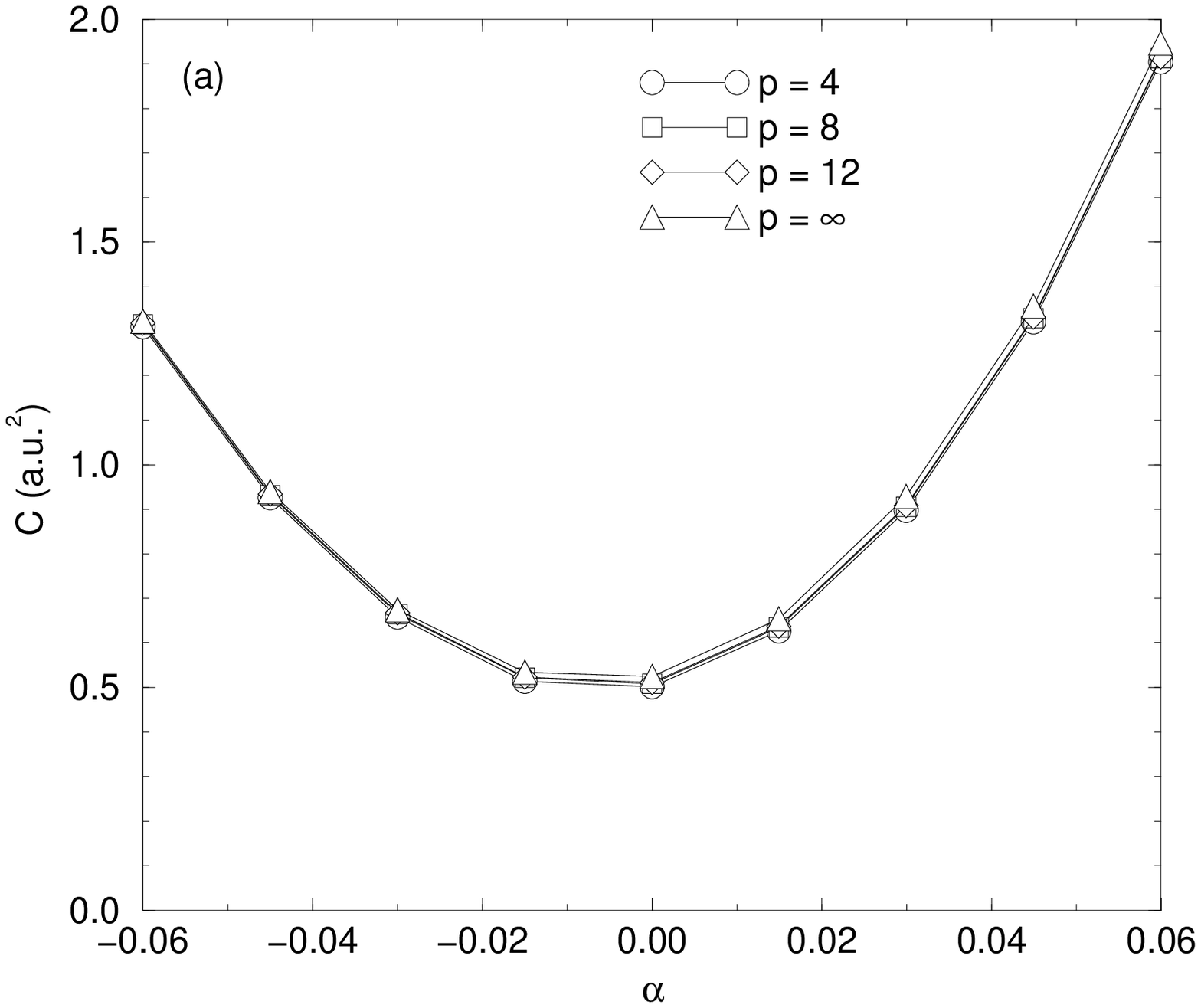}

\epsfysize=8cm \epsfbox{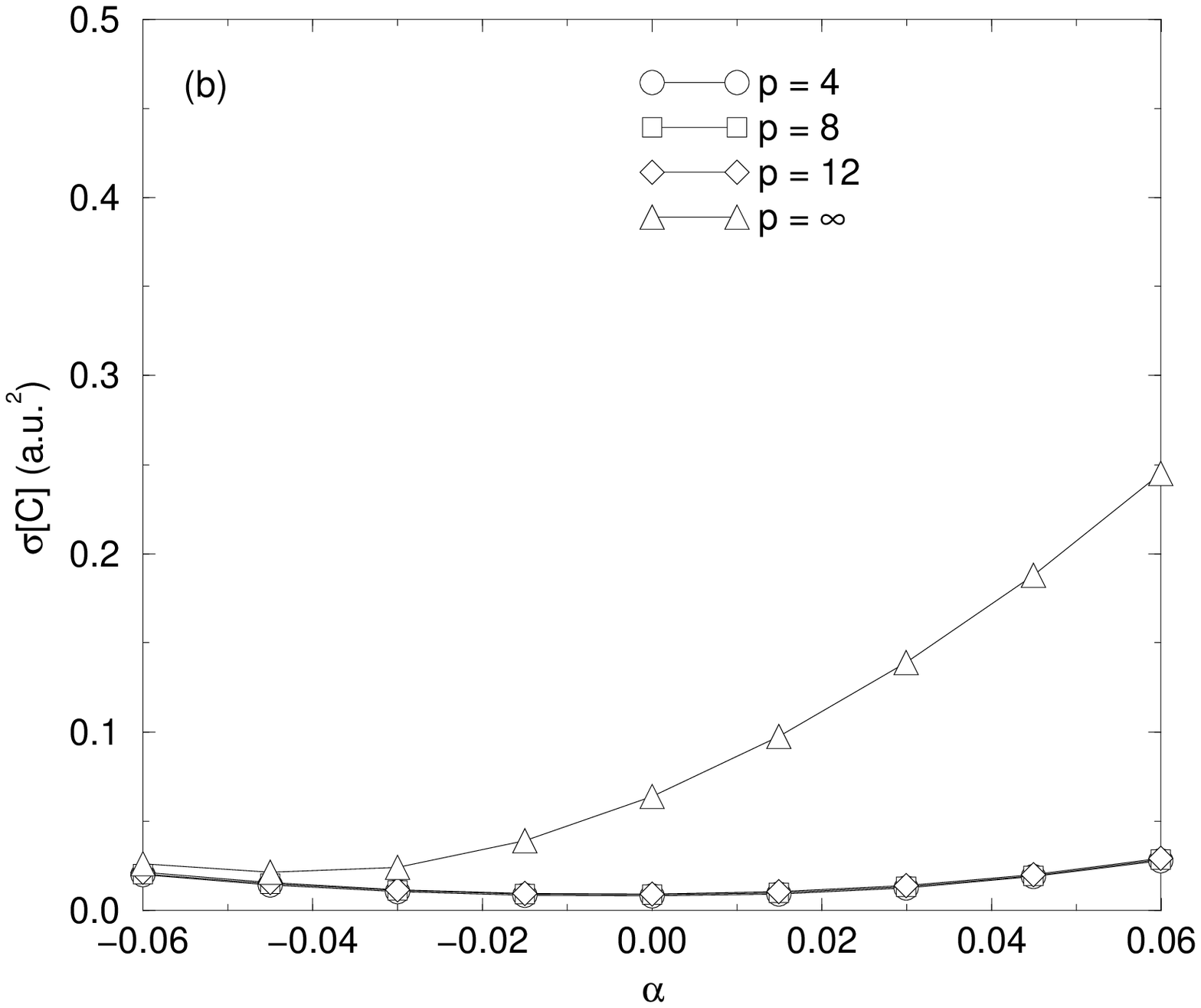}

\epsfysize=8cm \epsfbox{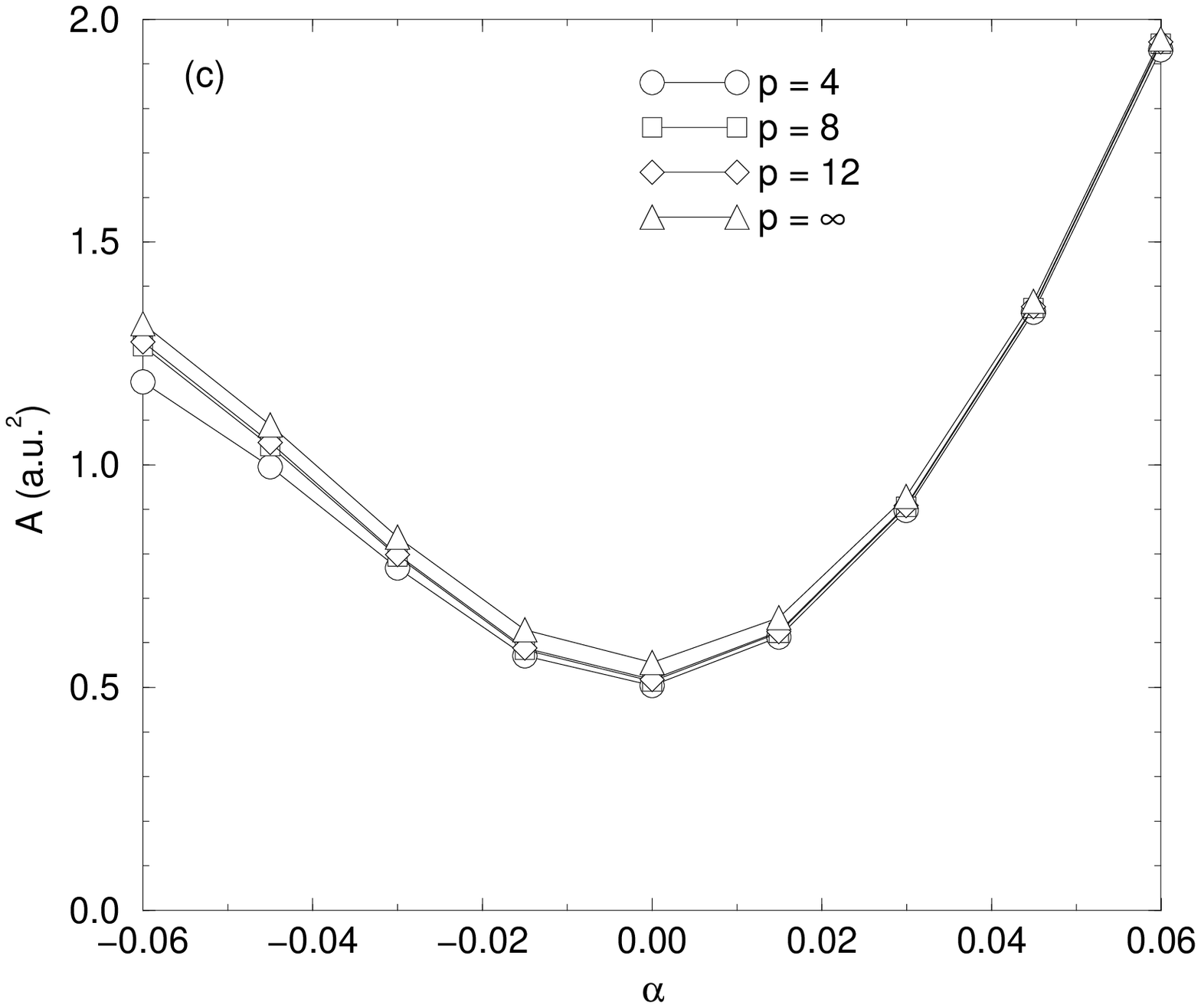}

\epsfysize=8cm \epsfbox{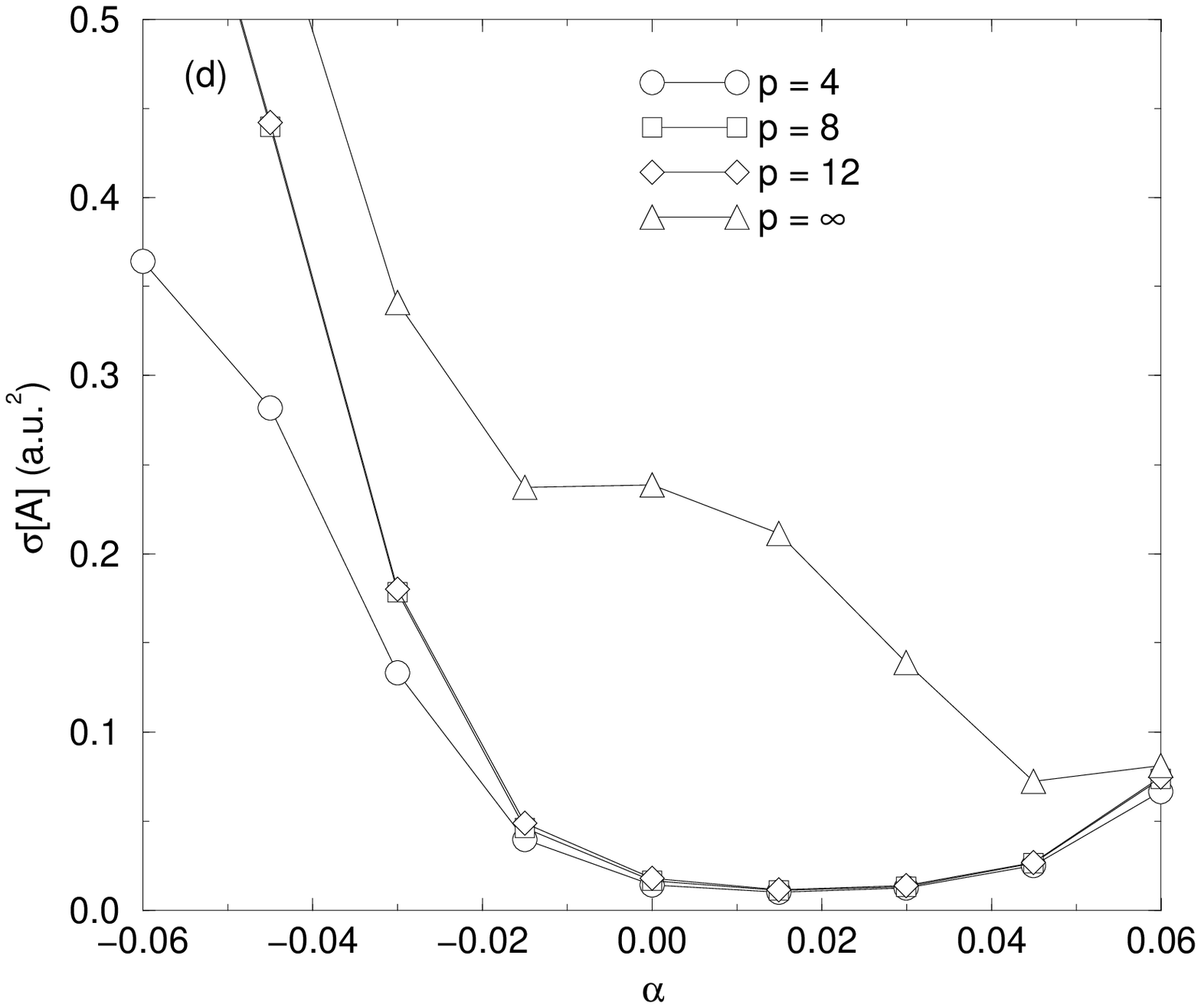}
\caption{The effect of limiting outlying energies on (a,b) objective
function $C$ (the unweighted variance) and (c,d) objective function
$A$ (the weighted variance) with $\alpha_0$=0.03. Outlying energies
are limited as in Eq.~\ref{Delta} with the values of $p$ shown.}
\label{fig4}
\end{figure}

\begin{figure}
\epsfysize=8cm \epsfbox{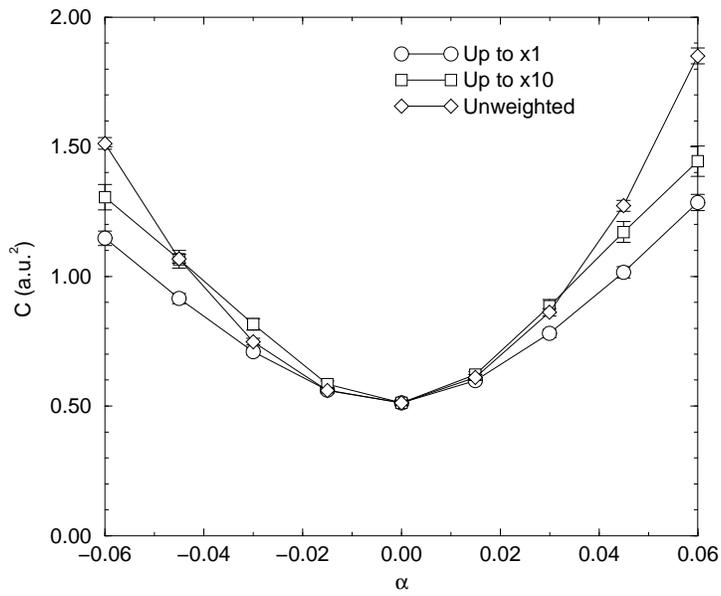}
\caption{The unweighted objective function $C$ generated with
$\alpha_0=0$ and with limiting of the weights.}
\label{fig5}
\end{figure}

\begin{figure}
\epsfysize=8cm \epsfbox{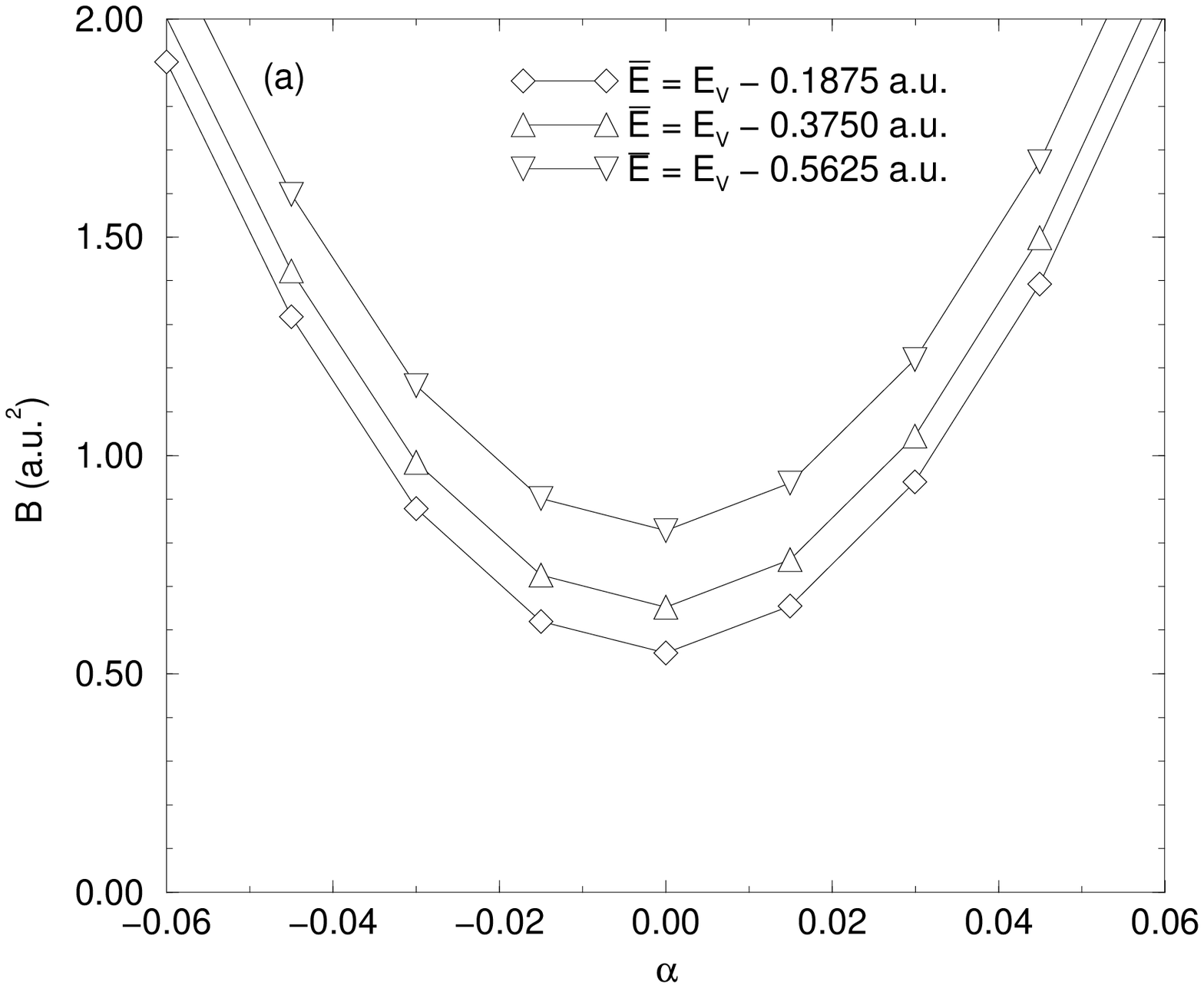}

\epsfysize=8cm \epsfbox{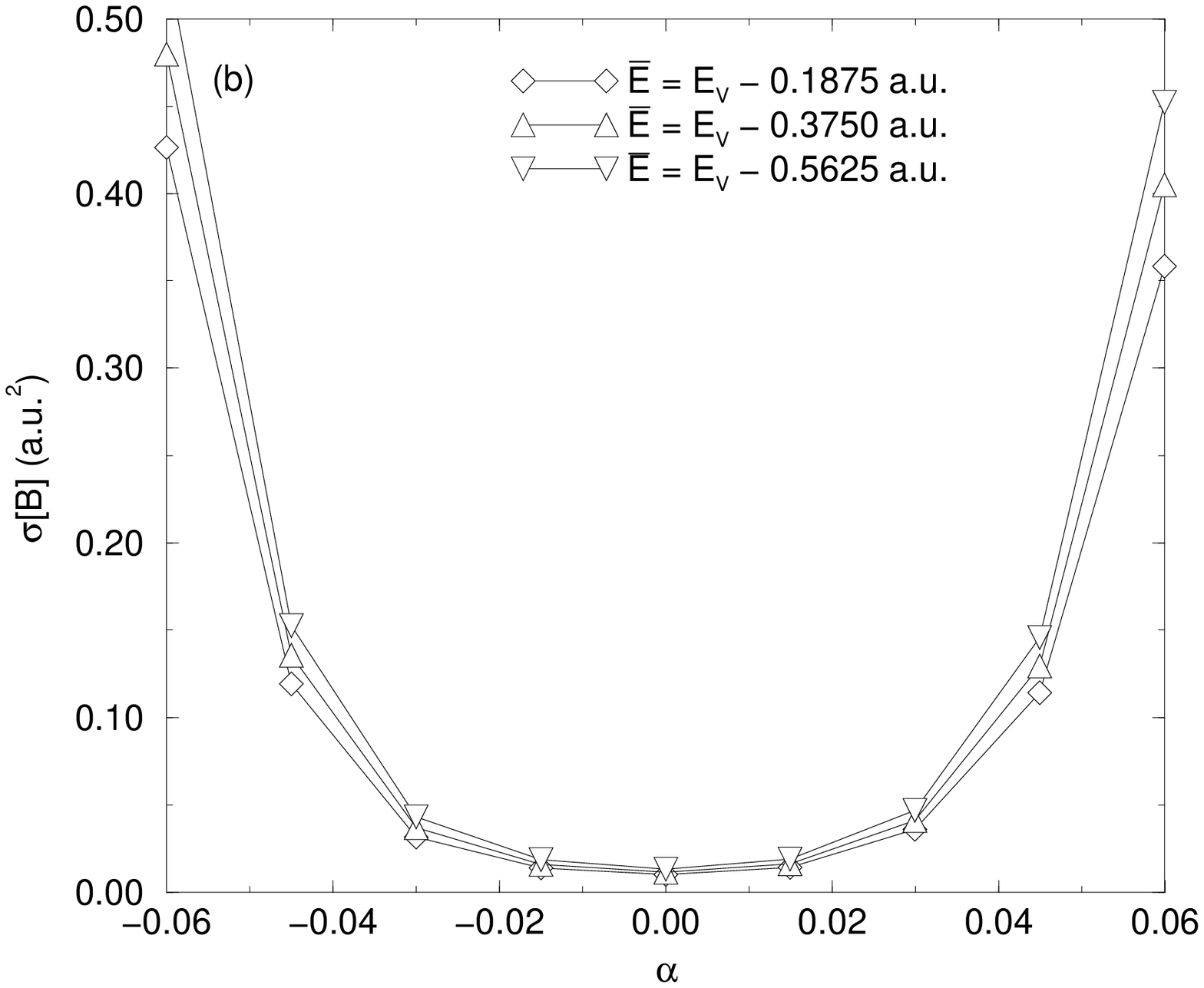}
\caption{Objective function $B$ versus $\alpha$ with $\alpha_0$ = 0
and $\bar{E} < E_{\rm V}$.}
\label{fig6}
\end{figure}

\begin{figure}
\epsfysize=8cm \epsfbox{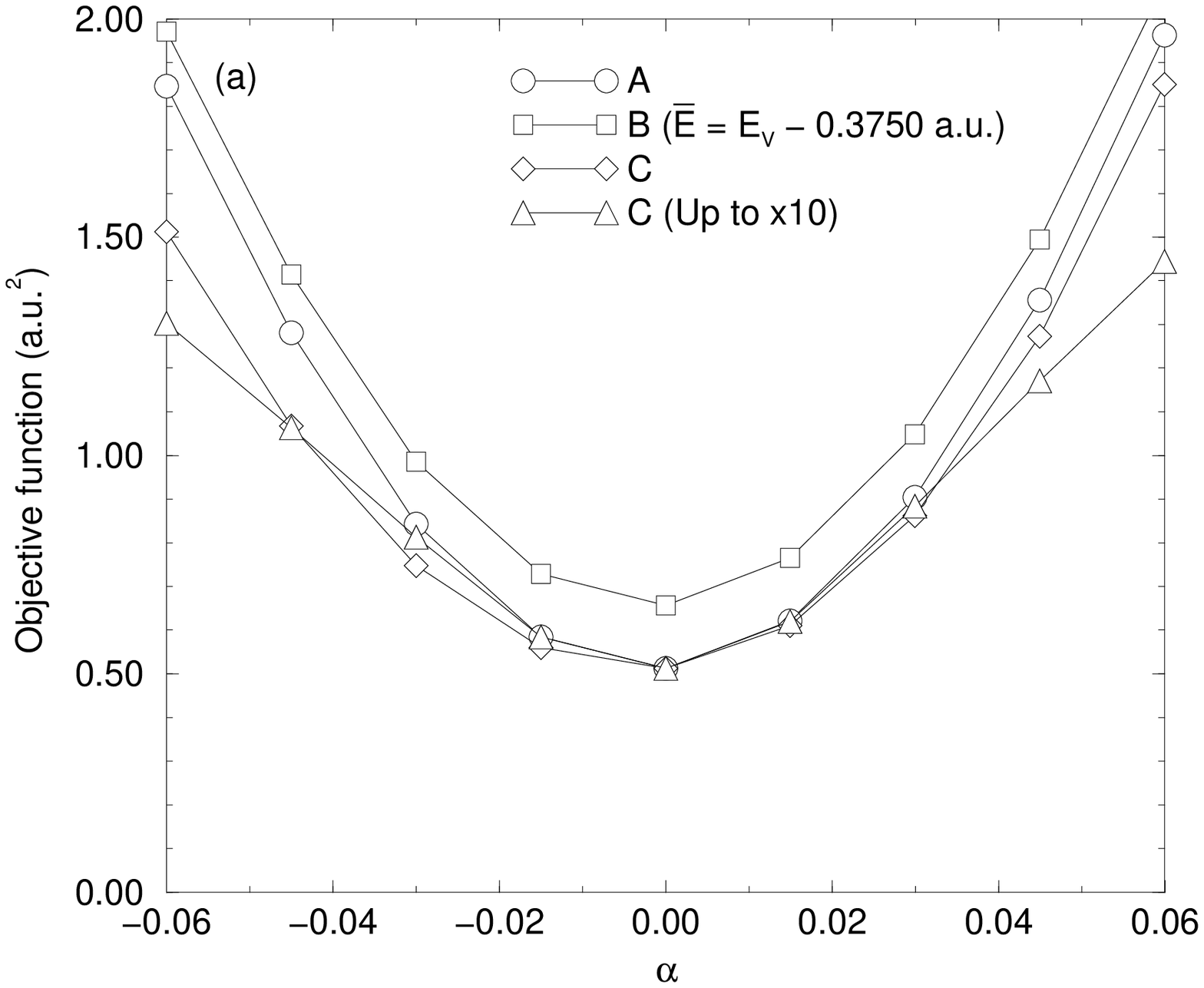}

\epsfysize=8cm \epsfbox{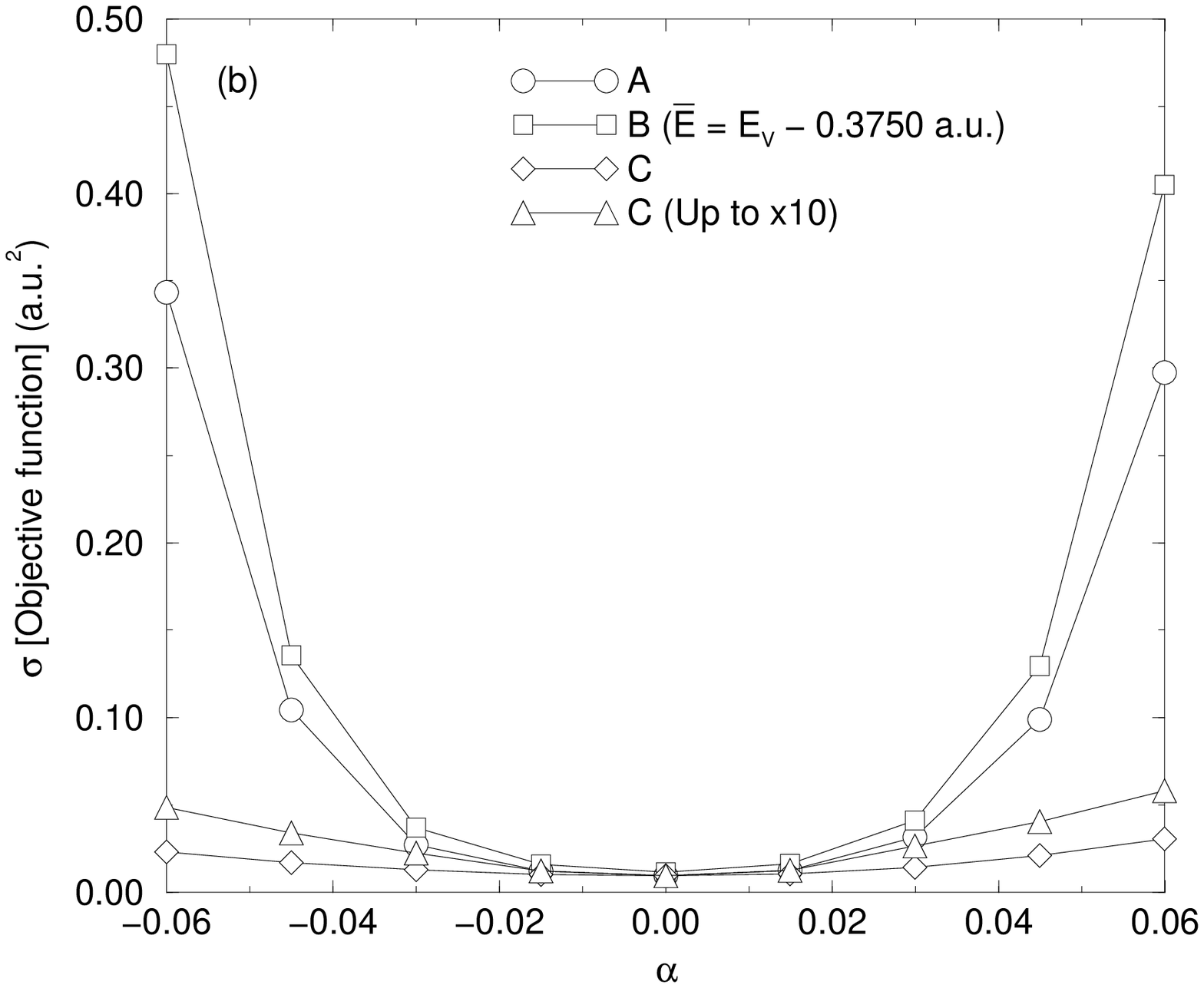}
\caption{Comparison of variance--like objective functions with
$\alpha_0$ = 0 as a function of $\alpha$.}
\label{fig7}
\end{figure}

\noindent $N_s$ is the total number of configurations and $p$ is
typically chosen to be 8, although varying $p$ from 4 to 12 makes no
significant difference to the results.  We include the factor of $N_s$
rather than limiting the energies beyond a given number of standard
deviations to incorporate the concept that as more configurations are
included, the sampling is improved.  In the limit of perfect sampling,
$N_s\rightarrow\infty$, the objective functions are unchanged.  For
our silicon system, the percentage of configurations having their
local energies limited by this procedure, with $p$ = 8, is only
0.024\% and 0.047\% for $\alpha_0$ = 0.03 and 0, respectively, which
corresponds to those beyond 5.7 standard deviations from the mean.
The effect of limiting the outlying local energies is illustrated in
Fig.~\ref{fig4}.  In Figs.~\ref{fig4}a and c we plot the mean values
of the objective functions $C$ and $A$ versus $\alpha$ for
configurations generated with $\alpha_0$ = 0.03, with values of the
limiting power, $p$, in Eq.~\ref{Delta}, of 4, 8, 12 and infinity (no
limiting), while in Figs.~\ref{fig4}b and d we plot their variances.
The mean values of $C$ are hardly affected by the limiting, while
those of $A$ are only slightly altered.  The smaller variances of $C$
and $A$ obtained by limiting the values of the local energy are very
clear.  In fact, if the local energies are not limited then the
variances of the objective function are not very accurately
determined, even with our large samples of $0.96\times 10^{6}$
configurations.  Similar results hold for configurations generated
with $\alpha_0$ = 0.  We are not aware of other workers limiting the
local energies in this way.  This method can significantly reduce the
variance of the variance--like objective functions without
significantly affecting their mean values.  Limiting the local
energies is even more advantageous when small numbers of
configurations are used.  As mentioned above, in practical
applications one evaluates the objective functions as averages over a
sample of some given size, so that the variance of interest is the
variance for that block size.  When the block size is small the
variances of objectives functions $A$ and $C$ increase, but this
effect is greatly reduced by limiting the local energies.  Limiting
the local energies in the way we have described gives significantly
better numerical behaviour for all the variance--like objective
functions and therefore all data shown in Figs.~\ref{fig5}--\ref{fig7}
have been limited with $p$=8, unless explicitly stated otherwise.

Limiting the values of the weights is a crucial part of the variance
minimization procedure for large systems.  Comparison of
Figs.~\ref{fig4}b and d shows that the variance of the unweighted
objective function $C(\alpha)$ is smaller than that of the weighted
objective function, $A(\alpha)$, for all values of $\alpha$, provided
one limits the local energies.  The variances close to the minimum are
similar but away from the minimum the variance of $A$ increases much
more rapidly than that of $C$.  The smaller variance of $C$ indicates
the superior numerical stability of the unweighted function.
Qualitatively similarly behaviour occurs for configurations generated
with $\alpha_0$ = 0.  A commonly used alternative to setting the
weights equal to unity is to limit the maximum value of the weights.
In Fig.~\ref{fig5} we show data for objective function $C$ with the
largest value of the weights limited to multiples of 1 and 10 times
the mean weight, along with data for the weights set to unity.  In
this graph the standard deviations of the objective functions are
plotted as error bars.  Fig.~\ref{fig5} shows that the variance of $C$
is reduced as the weights are more strongly limited, but the lowest
variance is obtained by setting the weights to unity.  In addition,
when the weights are limited the curvature of the objective function
is reduced, which makes it more difficult to locate the minimum.  We
therefore conclude that setting the weights to unity gives the best
numerical stability.

Finally, we study the effect of using the objective function
$B(\alpha)$ (Eq.~\ref{B}), in which the variational energy, $E_{\rm
V}$, is replaced by a fixed reference energy, $\bar{E}$, which is
chosen to be lower than the exact energy.  In Fig.~\ref{fig6} we show
the objective function $B(\alpha)$ versus $\alpha$ for configurations
generated with $\alpha_0$ = 0.  The overall shapes of the curves are
hardly changed as $\bar{E}$ is decreased, although the variance of the
objective function slowly increases.  If $\bar{E}$ is chosen to be too
low then a significant amount of energy minimization is included and
the numerical stability deteriorates.  The objective function $B$ does
have the property that its variance is independent of the block size,
so that it does not show the increase in variance at short block
sizes, but in practice we have not found this to be an important
advantage.  Using a value of $\bar{E}$ slightly below $E_{\rm V}$
appears to offer no significant advantages.

A direct comparison of the different variance--like objective
functions is made in Fig.~\ref{fig7}.  The behaviour of the following
objective functions are displayed: (i) $A$, (ii) $B$ with $\bar{E} =
E_{\rm V} -0.3750$ a.u., (iii) $C$, and (iv) a variant of $C$ with the
maximum value of the weights limited to 10 times the mean weight.
Limiting outlying values of the local energy improves the behaviour of
all the objective functions, so in each case we have limited them
according to Eq.~\ref{Delta} with $p$=8.  The mean values of the
objective functions are plotted in Fig.~\ref{fig7}a, which shows them
to behave similarly, with the positions of the minima being almost
indistinguishable.  However, the curve for the variant of $C$ with
limited weights is somewhat flatter, which is an undesirable feature.
The standard deviations of the objective functions are plotted in
Fig.~\ref{fig7}b, and here the differences are more pronounced.  The
unweighted variance, $C$, has the smallest variance, which is slightly
smaller than that of the variant of $C$ with strongly limited weights.
The variances of the objective functions which include the full
weights increase rapidly away from $\alpha$=0.  This rapid increase is
highly undesirable and can lead to numerical instabilities.

\section{Conclusions}
\label{conclusions}

We have analysed energy and variance minimization schemes for
optimizing many--body wave functions, where the integrals involved are
evaluated statistically.  We have suggested two reasons why variance
minimization techniques are numerically more stable than energy
minimization techniques:

\begin{enumerate}

\item In variance minimization it is allowable to limit the weights or
set them equal to unity, which reduces the variance of the objective
function while introducing only a ``weak bias'', which disappears as
the process converges.  Altering the weights in energy minimization
normally leads to a badly behaved objective function.

\item Variance minimization, with or without altering the weights,
shows greater numerical stability against errors introduced by finite
sampling because the positions of the minima of the variance are not
shifted by the finite sampling, whereas those of the (properly
weighted) energy are.

\end{enumerate}
 
We have studied optimization strategies for a realistic model of the
valence electronic structure of diamond--structure silicon.  The best
strategy we have found is:

\begin{enumerate}
\item Minimize the variance of the unweighted local energy (objective
function $C$, Eq.~\ref{C}).
\item Limit outlying values of the local energy according to
Eq.~\ref{Delta}.
\item Regenerate the configurations several times with the updated
parameter values until convergence is obtained.
\end{enumerate}

This stategy may be applied to both ground and excited states of
atoms, molecules and solids.  It has been designed to be optimal for
systems containing many electrons.  The behaviour that we have
observed in numerous wave function optimizations for large systems is
consistent with the analysis presented in this paper and indicates
that the above optimization strategy is robust, accurate and
efficient.

\section{Acknowledgments}
Financial support was provided by the Engineering and Physical
Sciences Research Council (UK).  Our calculations are performed on the
CRAY T3E at the Edinburgh Parallel Computing Centre, and the Hitachi
SR2201 located at the Cambridge HPCF.

\onecolumn

\clearpage

\clearpage

\begin{references}

\bibitem{vmc} W.~L.~McMillan, Phys. Rev. {\bf 138}, A442 (1965).

\bibitem{hammond} B.~L.~Hammond, W.~A.~Lester, Jr., and
P.~J.~Reynolds, {\em Monte Carlo Methods in ab initio quantum
Chemistry\/}, (World Scientific, Singapore, 1994).

\bibitem{dmc} D.~Ceperley, G.~Chester, and M.~Kalos, Phys. Rev. B {\bf
16}, 3081, (1977).

\bibitem{conroy} H.~Conroy, J. Chem. Phys. {\bf 41} 1331 (1964).

\bibitem{uww} C.~J.~Umrigar, K.~G.~Wilson, and J.~W.~Wilkins, Phys.
Rev. Lett. {\bf 60}, 1719 (1988).

\bibitem{umrigar} C.~J.~Umrigar, M.~P.~Nightingale, and K.~J.~Runge,
J. Chem. Phys. {\bf 99}, 2865 (1993).

\bibitem{filippi} C.~Filippi and C.~J.~Umrigar, J. Chem. Phys. {\bf
105} 213 (1996).

\bibitem{schmidt} K.~E.~Schmidt and J.~W.~Moskowitz,
J. Chem. Phys. {\bf 93} 4172 (1990).

\bibitem{varmin_96} A.~J.~Williamson, S.~D.~Kenny, G.~Rajagopal,
A.~J.~James, R.~J.~Needs, L.~M.~Fraser, W.~M.~C.~Foulkes, and
P.~Maccallum, Phys. Rev. B {\bf 53}, 9640 (1996).

\bibitem{barnett} R.~N.~Barnett, Z.~Sun, and W.~A.~Lester, Jr., Chem.
Phys. Lett. {\bf 273} 321 (1997).

\bibitem{finsize_98} P.~R.~C.~Kent, R.~Q.~Hood, A.~J.~Williamson,
R.~J.~Needs, W.~M.~C.~Foulkes, and G.~Rajagopal, unpublished.

\bibitem{fahy_opt} S.~Fahy, unpublished.

\bibitem{cohen} M.~L.~Cohen and T.~K.~Bergstresser, Phys. Rev. {\bf
141} 789 (1966).

\end{references}
\end{document}